# Magnetic properties of ZnO diluted by Mn through a Monte Carlo Investigation

L. B. Drissi[1, 2]

1- INANOTECH (Institut of Nonmaterials and Nanotechnology)- MASCIR
(Moroccan Foundation for Advanced Science, Innovation and Research)
Rabat, Morocco
2- International Centre for Theoretical Physics, ICTP, Trieste, Italy,

ldrissi@ictp.it, b.drissi@mascir.com

## Abstract

Diluted Magnetic Semiconductors (DMS) doped with a small concentration of magnetic impurities inducing ferromagnetic DMSs have attracted a lot of attention in the few last years. In particular, DMS based on III-V and II-VI semiconductors doped with transition metal are deeply investigated by both theoretical and experimental scientists for their promoting applications in spintronics. In this work, we present the magnetic properties of doped Mn ions in semi-conductor for different carrier's concentration. For the case of $Zn_{1-x}Mn_xO$, the results of our calculations, dealt within Monte Carlo study using a Heisenberg Hamiltonian based on the RKKY interaction, show well converged MC data based on this Hamiltonian for different carrier concentrations.

## Introduction

The study of the magnetic features of some p-type doped semiconductors such as ZnO, GaN, GaAs and ZnTe. by numerical simulation techniques has attracted great attention. For the case of ZnO, the investigations have led to several results. The ferromagnetic properties have been predicted using first-principles calculations and they were also proved experimentally. The incorporation of Mn into ZnO not only causes the introduction of magnetic moments but also increases the band gap. However, from experimental point of view, there is a controversy concerning the existence and the nature of magnetism in Mn-doped ZnO systems. Indeed, there exists some data where no ferromagnetism has been found. While there is also other data on many systems which exhibit ferromagnetism with Curie temperature higher than room temperature. Even if ZnO have been well investigated using ab-initio calculations, we are not aware of any studies for Mn-doped ZnO using Monte Carlo. This work, present data resulting from Monte Carlo simulations. More precisely, spontaneous magnetization, specific heat, magnetic susceptibility and the Curie temperature $T_c$ have been evaluated for different concentrations of magnetic impurities and carriers of Mn-doped ZnO.

# Basic data and Results

Recall that according to the Zener model approach, ferromagnetism in ZnO originates from the RKKY–like interaction between the localized Mn spins via the delocalized holes carriers' spins. So our system $Zn_{1-x}Mn_xO$ is described by Heisenberg Hamiltonian based on the RKKY interaction. The data got from Monte Carlo simulations (MC) for the series of $Zn_{1-x}Mn_xO$ studied lead to the results that we summarize in the following figures.

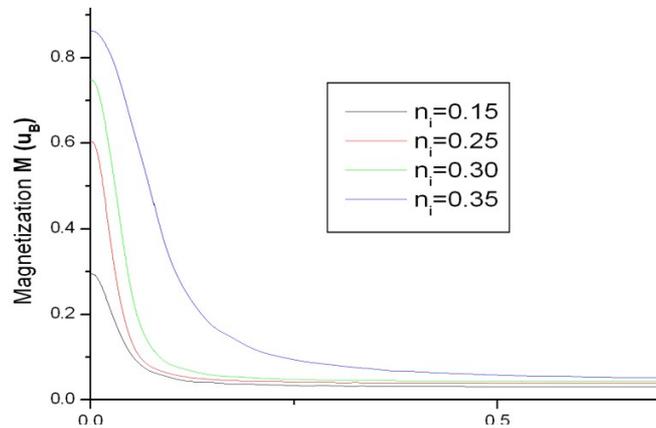

Figure 1: Magnetization M as function of temperature for different magnetic cations concentrations, $n_i$, and for a fixed carrier concentration $n_c$=0.18 in the case of short ranged RKKY interaction. When T increases the magnetization M decreases and goes to zero.

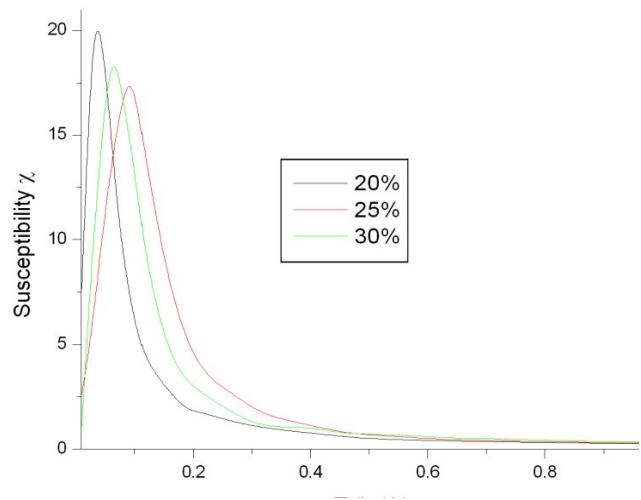

Figure 2: Temperature dependence of the susceptibility for a fixed carrier concentration $n_c$=0.18 and for $n_i$=0.2, 0.25 and 0.3. The three curves peak for different values of T that decreases as the concentration $n_i$ reduces. Notice that the peaks suggest that thermal fluctuation are driven phase transition from FM to paramagnetic.

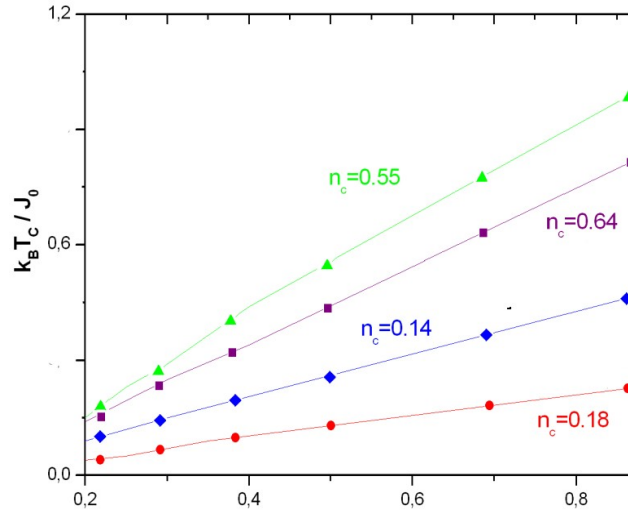

Figure 3: The variation of $T_C$ in function of the impurity concentrations $n_i$ in the large interval [0.2-0.9] for four different holes density values $n_c$, namely 0.14, 0.18 and 0.55, 0.64. $T_C$ increasing linearly with impurity concentration for the four values of $n_c$. Generally, this behavior of $T_C$ teaches us that hole density has also an effect.

## Conclusion

The Monte Carlo simulations, of $Zn_{1-x}Mn_xO$ with concentrations $n_i$ of Mn dopant ranging from 0.15 to 0.35, are performed to determine $T_c$ of those systems. This study teaches us also that $T_c$ is function of many parameters not only the magnetic cation concentrations $n_i$ but also it depends crucially on holes density $n_c$.